\documentclass{arxivsimilar}
\begin{document}
\title{Offline Multi-Agent Reinforcement Learning with a Physics-Informed World Model for Cooperative Mixed Traffic Control}

\author{\textbf{Lu Liu}\qquad \textbf{Chi Xie}\qquad \textbf{Xi Xiong}\thanks{Corresponding author.}\\
College of Transportation, Tongji University\\
Shanghai, China, 201804\\
\texttt{luliu0720@tongji.edu.cn}, \texttt{chi.xie@tongji.edu.cn},\\
\texttt{xi\_xiong@tongji.edu.cn}}
\date{}

\maketitle

% ---------- Abstract ----------
\begin{abstract}
This study investigates cooperative control of connected and automated vehicles (CAVs) at partially observable highway bottlenecks in mixed traffic, aiming to mitigate congestion without relying on complete global traffic states or online trial-and-error. We propose a physics-informed world model-based offline multi-agent reinforcement learning framework that reconstructs a physically interpretable global traffic state from local CAV observation-action histories, with coupled macroscopic-microscopic traffic dynamics providing physics-based supervision. A probabilistic ensemble world model learns traffic-state transitions and system rewards, while model disagreement quantifies epistemic uncertainty. Multi-step imagined rollouts with pessimistic rewards and uncertainty-driven truncation are then used for offline policy learning. Experiments in a SUMO-based on-ramp bottleneck using approximately $1\times10^6$ offline transitions show that physics supervision improves state reconstruction and world-model prediction accuracy.

\end{abstract}

 % \section{Introduction}
\section{Introduction}

With the rapid development of automated driving and vehicular networking technologies, connected and automated vehicles (CAVs) are evolving from passive traffic participants into active control agents. Unlike conventional traffic control strategies that rely on roadside infrastructure, CAVs can directly regulate surrounding driving behavior and local traffic evolution through their own control actions, providing a new approach for link-level traffic flow control \cite{LU202226partb}. This capability is particularly important for highway bottlenecks, such as on-ramp merging and lane-drop areas, where intensified vehicle interactions and capacity variations can amplify local speed disturbances, leading to queue formation and capacity degradation \cite{CHUNG200782PartB}. Although recent studies have explored cooperative CAV control to improve traffic efficiency, most of them assume that the global traffic state is available or that the observations of all CAVs can collectively cover the entire traffic system. This assumption is difficult to satisfy at highway bottlenecks, where CAVs are constrained by limited sensing and communication ranges and can only observe partial surrounding traffic conditions.

\begin{figure}[h]
    \raggedright
    \includegraphics[width=0.7\linewidth]{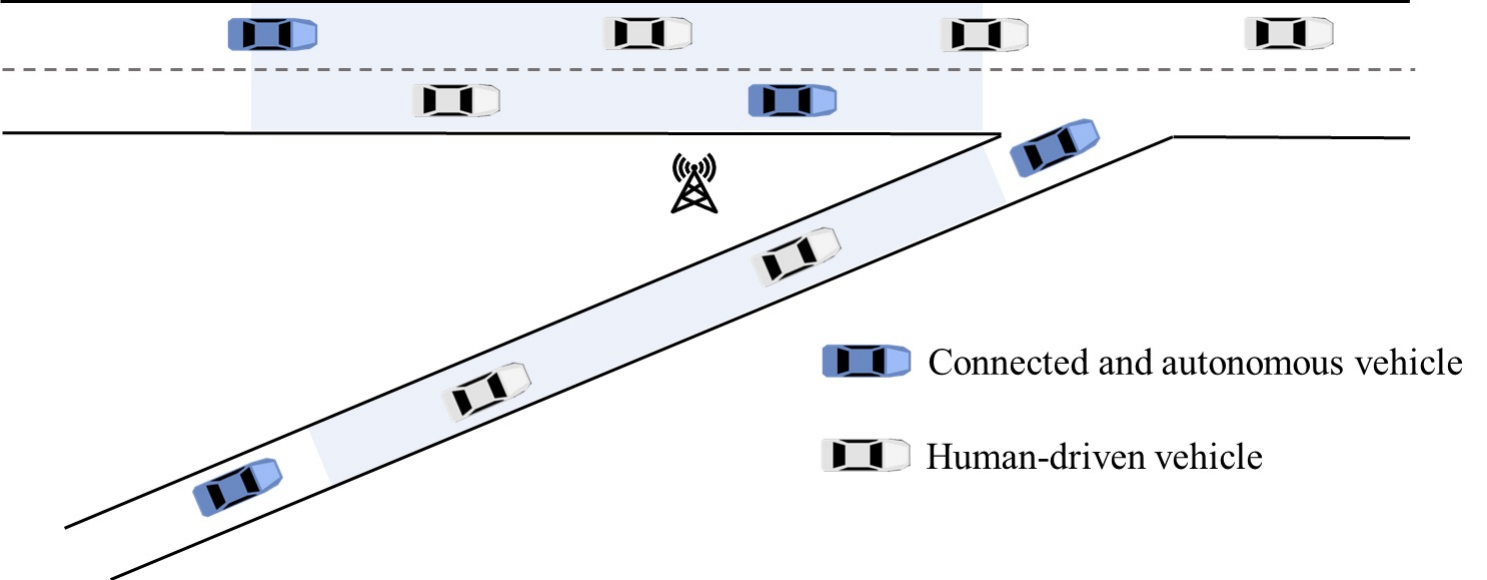}
    \caption{Mixed-traffic on-ramp bottleneck with the CAV control zone}
    \label{ramp}
\end{figure}

To enable effective coordination among multiple CAVs for mitigating traffic congestion, this paper proposes a physics-informed world model-based offline multi-agent reinforcement learning framework for mixed traffic at bottlenecks with limited global visibility.
As shown in Fig.~\ref{ramp}, we consider a typical on-ramp merging scenario where human-driven vehicles (HDVs) and CAVs coexist. The blue shaded area denotes the CAV control zone upstream of the merging point. Once entering this zone, each CAV obtains local traffic information within a limited range through onboard sensors and vehicle-to-everything (V2X) communication. Based on these partial observations and historical interactions, CAVs execute longitudinal acceleration control to actively regulate upstream traffic flow. In contrast to fully automated traffic, mixed traffic involves highly stochastic car-following and merging behaviors of HDVs, making the traffic response to CAV control difficult to predict \cite{Xiong2024ITS}. Therefore, online trial-and-error learning on real roads may introduce potential safety risks. To address this issue, this paper focuses on learning a world model from historical observation-action data of CAVs to capture bottleneck traffic evolution and support reliable CAVs cooperative control.

Early efforts in CAV control were dominated by rule-based approaches, which designed control policies based on vehicle spacing, relative speed, and leading-vehicle motion information \cite{Liao2022ITS}.
To extend CAV control from individual vehicle regulation to traffic-level coordination, many studies have combined macroscopic-microscopic coupled models with model predictive control \cite{QIU2023Cooperative}. 
However, the performance of these methods depends on the modeling accuracy of driving behaviors and traffic demand variations \cite{LI2020225}. Moreover, as the number of CAVs increases, the optimization scale grows rapidly, leading to substantial communication and computational burdens \cite{Feng2021}.

Data-driven methods have therefore attracted increasing attention, among which reinforcement learning (RL) has shown strong potential for cooperative CAV control by learning from interaction feedback \cite{WANG2023359, li2023developing}.
Online policy exploration in real traffic systems may induce unsafe driving behaviors and incur prohibitive costs \cite{cang2021behavioral}. Offline RL, which learns from pre-collected datasets without continual environment interaction during training, has therefore emerged as a promising direction. Nevertheless, offline RL suffers from distribution shift, as poorly covered states or actions may lead to suboptimal or unsafe decisions \cite{pmlr-v202-ran23a}. For example, Fang et al. deployed offline RL using real-world driving data and found that limited data coverage led to unsatisfactory CAV control performance \cite{Fang2022ITSC}. Although policy constraints and value regularization have been widely used to improve policy reliability \cite{huang2025uncertainty}, CAVs coordination introduces additional challenges: even when each agent selects an in-distribution action, the resulting joint action may still be out of distribution. Moreover, without interactive feedback, effective multi-agent coordination is difficult to guarantee \cite{Zhou2025aaai}. As a result, offline RL has seen limited application in CAV control.

World model-based offline RL offers a promising way to address these limitations. By learning environment dynamics and generating imagined experience, these methods can improve sample efficiency and enhance lookahead prediction capability \cite{Ross2021ICML}. To mitigate model exploitation in offline settings, existing studies often incorporate uncertainty estimation, pessimistic rewards, or penalties for uncertain regions \cite{wiesemann2013robust}. 
However, most world model methods are designed for general control tasks or single-vehicle decision-making and typically rely on relatively complete environment states \cite{Kang2025ICLR}. Such an assumption is often difficult to satisfy in practice because of limited sensing and communication capabilities.
Therefore, some studies directly concatenate the local observations of multiple CAVs to construct an approximate global state for centralized training \cite{rauch2024cooperative}. However, such representations may introduce substantial information redundancy, and their dimensionality increases with the number of CAVs. Other studies learn low-dimensional latent states to obtain more compact system representations \cite{hafner2019dream}. Although latent representations can reduce the modeling complexity, they typically lack explicit traffic-physical semantics, potentially leading to predictions that are inconsistent with actual traffic evolution.
In addition, multi-agent systems involve larger joint state-action spaces, making it difficult for finite offline datasets to sufficiently cover the state-action regions that may be visited by the learned policy \cite{bui2024comadice}. Under such distribution shifts, the policy may exploit prediction errors of the world model in poorly covered regions, thereby generating unreliable imagined experiences \cite{Kidambi2020NIPS}.

Based on the above analysis, applying world models to cooperative multi-CAV control in partially observable mixed traffic still faces three interrelated challenges: how to construct an effective global state that captures overall traffic evolution from distributed local observations, how to improve the consistency of the learned world model with traffic-flow physics, and how to mitigate the impact of model prediction errors on imagined policy learning under limited offline data coverage.

This paper proposes a physics-informed world model-based offline multi-agent reinforcement learning framework for mixed-traffic bottlenecks. 
First, temporal traffic information is extracted from the local observation-action histories of multiple CAVs, and a structured global traffic state consisting of traffic density, average speed, and microscopic CAV states is reconstructed through traffic-cell-based feature aggregation. 
Unlike directly concatenating local observations or learning latent states without explicit physical semantics, the proposed method incorporates a coupled macroscopic-microscopic traffic model to provide physics-based consistency supervision for state reconstruction, enabling the reconstructed state to explain local observations while better reflecting traffic-flow evolution. Based on this reconstructed state, a probabilistic ensemble world model is developed to learn traffic-state transitions and system rewards under joint CAV control, while epistemic uncertainty is quantified through prediction disagreement among ensemble members. Multi-step imagined rollouts are then performed in the learned traffic world model, together with uncertainty-aware pessimistic rewards and rollout truncation, to reduce the risk of exploiting unreliable model predictions in poorly covered regions. 
Finally, cooperative multi-CAV policies are learned under the centralized training with decentralized execution (CTDE) paradigm, allowing each CAV to make decisions based solely on its own local observation history during execution, without online trial-and-error or access to the true global traffic state. 
A mixed-traffic on-ramp scenario is constructed using simulation of urban mobility (SUMO) to evaluate the proposed method in terms of world-model prediction accuracy, cooperative control performance, and sensitivity to different traffic demands and CAV penetration rates.

The remainder of this paper is organized as follows. Section~\ref{sec_preliminaries} introduces the coupled macroscopic-microscopic traffic model and the Dec-POMDP formulation. Section~\ref{sec_analysis}  presents the proposed physics-informed world model-based offline multi-agent reinforcement learning method. Section~\ref{sec_results} describes the experimental setup and analyzes the results. Finally, Section~\ref{sec_conclude} concludes the paper.

 %3p
 \section{Preliminaries}
\label{sec_preliminaries}

\subsection{Decentralized POMDP Formulation}

The cooperative control problem of multiple CAVs in mixed-autonomy traffic is formulated as a decentralized partially observable Markov decision process (Dec-POMDP) \cite{bernstein2002complexity}, represented by the tuple $\langle \mathcal{N}, S, A, P, R, \Omega, O, \gamma \rangle$. 
$\mathcal{N}=\{1,2,\ldots,N\}$ denotes the set of CAVs operating within the coordination region. 
The road segment within the coordination region is partitioned into $M$ homogeneous cells with length $\Delta x$, indexed by $\mathcal\{J\}=\{1,2,\ldots,M\}$. 
The global state space is denoted by $S$, and at time $t$, the system state  $s(t)\in S$ characterizes the underlying traffic state of the mixed-autonomy system. Specifically, the global state is defined as $s(t)=\left(\rho(t),\bar v(t),e(t)\right)$,
where $\rho(t)=\{\rho_1(t),\rho_2(t),\ldots,\rho_M(t)\}$ and $\bar v(t)=\{\bar v_1(t),\bar v_2(t),\ldots,\bar v_M(t)\}$ denote the traffic density and average speed of the $M$ road cells, respectively. The term $e(t)$ contains the microscopic states of all CAVs, including their positions and velocities.

The joint action space is defined as $A=\prod_{i\in\mathcal{N}}A_i$, where $A_i$ is the action space of CAV $i$. In this study, the action of each CAV is represented by a continuous longitudinal acceleration command $a_i(t)\in A_i=[a_{\min},a_{\max}]$, where $a_{\min}$ and $a_{\max}$ denote the minimum and maximum admissible accelerations. The actions of all CAVs jointly form the action vector $a(t)=\{a_i(t)\}_{i\in\mathcal{N}}$.
The stochastic evolution of the traffic system is characterized by the transition function $P:S\times A\times S\rightarrow[0,1]$, which captures the combined effects of traffic-flow propagation, CAV control actions, and stochastic human-driving behaviors.

Let $\Omega=\prod_{i\in\mathcal{N}}\Omega_i$ denote the joint observation space, and let $O=\{O_i\}_{i\in\mathcal{N}}$ denote the set of local observation functions. For each CAV $i$, the observation function maps the global traffic state to its local observation, $o_i(t)=O_i(s(t))$.
The local observation of CAV $i$ is defined as $o_i(t)=\langle x_i(t),v_i(t),\varepsilon_i^{\mathrm{surr}}(t)\rangle$, where $x_i(t)$ and $v_i(t)$ denote the position and velocity of CAV $i$, respectively, and $\varepsilon_i^{\mathrm{surr}}(t)$ represents the surrounding traffic information perceived within its sensing or communication range. The observations of all CAVs constitute the joint observation vector $o(t)=\{o_i(t)\}_{i\in\mathcal{N}}$. The parameter $\gamma\in[0,1)$ is the discount factor.

All CAVs share a common team reward function $R:S\times A\rightarrow\mathbb{R}$. The immediate reward is designed to improve system-level traffic efficiency while suppressing excessive control actions,
\begin{align}
    r(t)=R(s(t),a(t))= w_1\sum_{j=1}^{M}\rho_j(t)\bar v_j(t) - w_2\sum_{i\in\mathcal{N}} a_i^2(t),
\end{align}
where the first term approximates the aggregated traffic flow over all road cells, and the second term penalizes aggressive acceleration or deceleration. The parameters $w_1>0$ and $w_2>0$ balance traffic efficiency and driving smoothness.

At each decision step $t$, CAV $i$ does not have access to the complete global state $s(t)$. Instead, it selects its control action based on its local interaction history $\tau_i(t)=(o_i(t-H:t),a_i(t-H:t-1))$, according to a decentralized policy, $a_i(t)\sim \pi_i(\cdot \mid \tau_i(t))$.
After the joint action $a(t)$ is executed, the traffic system evolves to the next state $s(t+1)$ according to $P(s(t+1)\mid s(t),a(t))$, and a team reward $r(t)=R(s(t),a(t))$ is generated. The objective is to learn the optimal decentralized policies $\pi^*=\{\pi_i^*\}_{i\in\mathcal N}$ that maximize the expected cumulative discounted return,
\begin{align}
    \pi^*=\arg\max_{\pi}\mathbb{E}_{\pi}\left[\sum_{t=0}^{\infty}\gamma^t r(t)\right].
\end{align}

\subsection{Traffic Flow Dynamics Prior}

To ensure the physical consistency of state inference and traffic evolution prediction, a macroscopic-microscopic coupled traffic flow model is introduced as a physics-informed prior.

At the macroscopic level, the mainstream traffic dynamics are described by the Lighthill-Whitham-Richards (LWR) conservation law \cite{Richards1956ShockWO},
\begin{align}
    & \partial_t \rho(x,t)+\partial_x q(x,t)=0,\\
    & q(x,t)=Q(\rho(x,t))=\rho(x,t)v(x,t),
\end{align}
where $\rho(x,t)$, $v(x,t)$, and $q(x,t)$ denote the traffic density, traffic speed, and traffic flow at location $x$ and time $t$, respectively. The function $Q(\rho)$ denotes the traffic flow fundamental diagram. In this study, a triangular fundamental diagram \cite{newell1993simplified} is adopted to characterize the relationship between traffic density and flow. The corresponding equilibrium speed is given by,
\begin{align}
    v(x,t)=V(\rho(x,t))=\frac{Q(\rho(x,t))}{\rho(x,t)}.
\end{align}

For computational implementation, the continuous traffic-flow model is discretized over the cell partition $\mathcal{J}$ using the cell transmission model (CTM) \cite{DAGANZO1994269}. For each cell $j\in\mathcal{J}$, the density evolution is given by,
\begin{align}
    \rho_j(t+1)= \rho_j(t) + \frac{\Delta t}{\Delta x} \left(q_{j-1}(t)-q_j(t) +r_j^{\mathrm{on}}(t)-r_j^{\mathrm{off}}(t)\right),
\end{align}
where $q_{j-1}(t)$ and $q_j(t)$ denote the upstream inflow and downstream outflow of cell $j$, respectively. The terms $r_j^{\mathrm{on}}(t)$ and $r_j^{\mathrm{off}}(t)$ represent the on-ramp inflow and off-ramp outflow.
The boundary flow between two adjacent cells is determined by the demand and supply functions \cite{JIN20121000PartB},
\begin{align}
    & q_j(t)=\min\{D_j(t),S_{j+1}(t)\},\\
    & D_j(t)=\min\{V_f\rho_j(t),q_{\max}\},\\
    & S_{j+1}(t)=\min\{w(\rho_{\max}-\rho_{j+1}(t)),q_{\max}\},
\end{align}
where, $V_f$ denotes the free-flow speed, $w$ denotes the backward wave speed, $\rho_{\max}$ denotes the jam density, and $q_{\max}$ denotes the maximum flow capacity.

Accordingly, the cell physical speed prior is given by
\begin{align}
    \bar v_j(t+1)=V(\rho_j(t+1)).
\end{align}

At the microscopic level, the motion of each CAV is described by a discrete-time kinematic model,
\begin{align}
    & v_i(t+1) = \mathrm{clip}\left(v_i(t)+a_i(t)\Delta t, 0, V_{\max}\right),\\
    & x_i(t+1) = x_i(t)+v_i(t)\Delta t,
\end{align}
where $x_i(t)$, $v_i(t)$, and $a_i(t)$ denote the position, velocity, and acceleration command of CAV $i$, respectively. The operator $\mathrm{clip}(\cdot)$ ensures that the updated speed remains within the admissible speed range.

To capture the impact of CAV control on local traffic flow dynamics, CAVs are modeled as controllable moving disturbances embedded in the macroscopic traffic stream. Their longitudinal control actions affect local car-following behavior, gap formation, and traffic flow propagation. Accordingly, a CAV-induced flow correction term is introduced into the CTM boundary flow,
\begin{align}
q_j(t)=\min\{D_j(t),S_{j+1}(t)\}+\Delta q_j^{\mathrm{CAV}}(t),
\end{align}
where $\Delta q_j^{\mathrm{CAV}}(t)$ represents the aggregate impact of nearby CAVs. A positive value of $\Delta q_j^{\mathrm{CAV}}(t)$ indicates an improvement in local throughput induced by cooperative CAV control, whereas a negative value indicates reduced local traffic throughput caused by traffic disturbances. The flow correction term is computed according to the moving-bottleneck Riemann formulation in the literature~\cite{Mladen2018Traffic}.

By combining the macroscopic traffic evolution and microscopic CAV dynamics, the physics-based state transition can be compactly written as
\begin{align}
\tilde{s}(t+1)=F_{\mathrm{phy}}(s(t),a(t)).
\end{align}
This physics-based transition serves as a structural prior rather than a complete replacement for the learned world model. It provides physical guidance for global state inference and future traffic evolution prediction, ensuring that the learned model remains consistent with traffic conservation principles.
 %4p
 \section{Methodology}
\label{sec_analysis}
\subsection{Framework Overview}

\begin{figure}[H]
    \raggedright
    \includegraphics[width=1.0\textwidth]{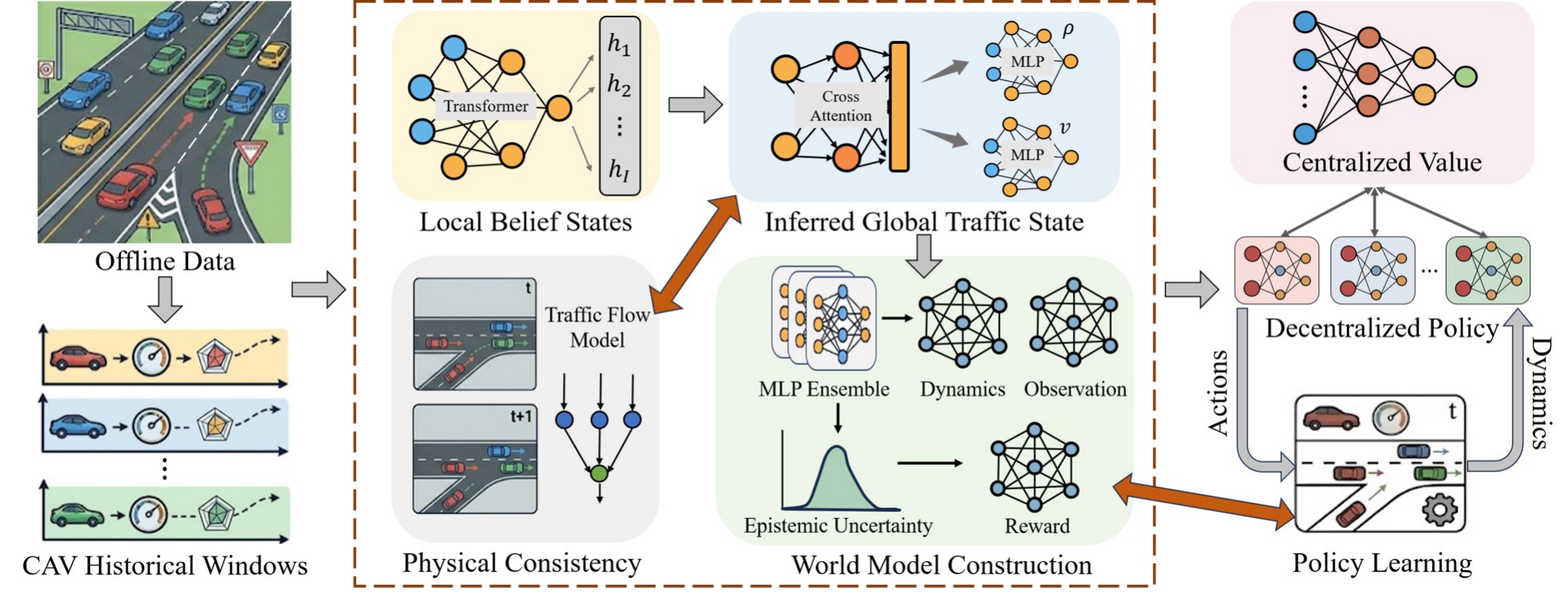}
    \caption{A traffic world model learning framework integrating traffic flow physics priors and epistemic uncertainty}
    \label{framework}
\end{figure}

In this section, we propose a physics-informed world model-based offline multi-agent reinforcement learning framework for cooperative control of multiple CAVs in mixed traffic. 
As illustrated in Fig.~\ref{framework}, the proposed framework consists of two main stages.

In the first stage, a traffic world model is trained using offline interaction data. By integrating traffic flow physics priors and epistemic uncertainty, the learned world model recovers physically meaningful global traffic states from local observations and predicts both traffic state evolution and system-level rewards.
In the second stage, world model-based imagination rollout and policy learning are conducted. Starting from historical states sampled from the offline dataset, the policy performs multi-step virtual interactions within the learned world model to generate additional imagined trajectories. These trajectories are then used to optimize cooperative multi-agent control policies under the centralized training with decentralized execution (CTDE) paradigm.

\subsection{Physics-Informed Global Traffic State Inference}

In this subsection, we develop a physically meaningful global traffic state inference module that recovers global traffic states from multi-vehicle historical observations and actions, thereby providing a state foundation for world model learning and centralized policy training.

For each CAV $i$ located within the cooperative control region at time $t$, the local interaction history $\tau_i(t)$ is used to characterize its recent motion, surrounding traffic evolution, and control response information. Then, a Transformer-based history encoder $\mathrm{Enc}_{\phi}$ is employed to extract the local belief representation of each CAV,
\begin{align}
    h_i(t)=\mathrm{Enc}_{\phi}\left(\tau_i(t)\right),
\end{align}
where $\phi$ denotes the learnable parameters. In this paper, all CAVs share the same encoder parameters, which enables the model to accommodate dynamic traffic scenarios where vehicles may enter or leave the control region and improves its generalization capability under different traffic demands and CAV penetration rates.

By aggregating the local beliefs of all CAVs, the proposed module further recovers a global traffic representation according to the spatial structure of the road. Specifically, each road cell is regarded as a spatial query unit. For each cell $j \in \mathcal{J}$, a cross-attention mechanism is used to adaptively aggregate the information relevant to that cell from the local beliefs of all CAVs,
\begin{align}
   & \alpha_{i,j}(t)=\frac{\exp\left(\frac{(\eta_j W_Q)(h_i(t)W_K)^{\top}}{\sqrt{d}}\right)}{\sum_{i'=1}^{N}\exp\left(\frac{(\eta_j W_Q)(h_{i'}(t)W_K)^{\top}}{\sqrt{d}}\right)},\\
  &  z_j(t)=\sum_{i=1}^{N}\alpha_{i,j}(t)\left(h_i(t)W_V\right),
\end{align}
where $\eta_j$ is the learnable spatial query vector associated with cell $j$, and $W_Q$, $W_K$, and $W_V$ are learnable projection matrices. The attention weight $\alpha_{i,j}(t)$ quantifies the degree to which cell $j$ extracts information from CAV $i$, while $z_j(t)$ represents the inferred state of cell $j$. The global traffic representation is denoted as $Z(t)=\left[z_1(t),\ldots,z_M(t)\right]$.

To enable the world model to learn traffic evolution dynamics in a state space that is consistent with traffic flow mechanisms, we further employ multilayer perceptrons (MLPs) to map the traffic representations into macroscopic traffic variables with clear physical meanings. These variables include cell density and average speed, which are essential for characterizing congestion propagation and flow variations,
\begin{align}
     \hat{\rho}_j(t)=\mathrm{MLP}_{\rho}\left(z_j(t)\right),\quad
     \hat{v}_j(t)=\mathrm{MLP}_{v}\left(z_j(t)\right).
\end{align}
The recovered density and average speed over the entire road segment are represented as $\hat{\rho}(t)=\left(\hat{\rho}_1(t),\ldots,\hat{\rho}_M(t)\right)$ and $\hat{v}(t)=\left(\hat{v}_1(t),\ldots,\hat{v}_M(t)\right)$, respectively. The density prediction head uses a Softplus activation function to ensure non-negativity, while the speed prediction head uses a Sigmoid mapping scaled to the free-flow speed range $[0,V]$.
In addition to macroscopic traffic states, we extract the individual state representation $e_i(t)$ of each CAV from its local observation $o_i(t)$, and denote the individual states of all CAVs as $e(t)=\left(e_1(t),\ldots,e_N(t)\right)$.
The recovered global traffic state is then constructed as $\hat{s}(t)=\left[\hat{\rho}(t),\hat{v}(t),e(t)\right]$.

To ensure the validity and interpretability of the inference module, an observation reconstruction constraint is introduced. Specifically, an observation decoder $O_{\psi}$ is constructed to map the recovered global state back to the corresponding local observation of each CAV $i$,
\begin{align}
    \hat{o}_i(t)=O_{\psi}\left(\hat{s}(t),i\right),
\end{align}
where $\psi$ denotes the parameters of the observation decoder. The observation reconstruction loss is defined as,
\begin{align}
    \mathcal{L}_{\mathrm{obs}}=\sum_{i=1}^{N}\left\|\hat{o}_i(t)-o_i(t)\right\|^2,
\end{align}
where $o_i(t)$ is the truth observation from the offline dataset. This loss encourages the recovered global state to explain the local observations of all CAVs. Moreover, after training, the observation decoder is used in the subsequent imagination rollout process, providing inputs for decentralized policy execution.

Relying solely on observation reconstruction may allow the recovered state to numerically explain local observations, but it does not necessarily guarantee consistency with real traffic flow evolution. Therefore, a traffic flow physics model is introduced as a soft constraint. Given the recovered state $\hat{s}(t)$ and the joint control action $a(t)$, the traffic flow physics model $F_{\mathrm{phy}}(\cdot)$ provides a physics-based reference state for the next time step,
\begin{align}
    s^{\mathrm{phy}}(t+1)=F_{\mathrm{phy}}\left(\hat{s}(t),a(t)\right).
\end{align}
The physical consistency loss is defined as,
\begin{align}
    \mathcal{L}_{\mathrm{phy}}=\left\|\hat{s}(t+1)-s^{\mathrm{phy}}(t+1)\right\|^2.
\end{align}
It should be noted that the physics model is not treated as the state transition function in this paper. Instead, it is used as a traffic flow prior to regularize the state inference process. This design constrains the learned state within a physically reasonable range defined by traffic conservation laws and fundamental diagram relationships, thereby improving the interpretability of state recovery and the stability of the constructed world model.

Overall, the training objective of the global traffic state inference module is formulated as,
\begin{align}
    \mathcal{L}_{\mathrm{state}}=\lambda_1\mathcal{L}_{\mathrm{obs}}+\lambda_2\mathcal{L}_{\mathrm{phy}},
\end{align}
where $\lambda_1>0$ and $\lambda_2>0$ are the weighting coefficients.

\subsection{Probabilistic Ensemble Traffic World Model}

This subsection focuses on learning the evolution dynamics of the traffic system in the state–action space. In mixed traffic, future traffic states are uncertain due to the stochastic behavior of human-driven vehicles (HDVs) and the control actions of CAVs.
Meanwhile, offline datasets cannot cover all possible state–action combinations, and a single predictive model may suffer from large errors in sparsely sampled regions. To address these issues, we develop a probabilistic ensemble traffic world model to jointly predict state transitions, estimate system-level rewards, and quantify epistemic uncertainty.

Specifically, we initialize $K$ MLPs with the same network architecture but independent parameters, denoted as ${\mathcal{M}{\theta_k}}{k=1}^{K}$. Each MLP $k$ takes the recovered global traffic state $\hat{s}(t)$ and the joint action $a(t)$ as inputs, and predicts the distribution of the next global traffic state,
\begin{align}
    \hat{s}(t+1)\sim\mathcal{N}\left(\mu_{\theta_k}\left(\hat{s}(t),a(t)\right),\operatorname{diag}\left(\sigma_{\theta_k}^{2}\left(\hat{s}(t),a(t)\right)\right)\right),
\end{align}
where $\mu_{\theta_k}\left(\hat{s}(t),a(t)\right)\in\mathbb{R}^{d_s}$ and $\sigma_{\theta_k}\left(\hat{s}(t),a(t)\right)\in\mathbb{R}_{>0}^{d_s}$ denote the mean vector and standard deviation vector of the predicted state transition distribution, respectively. $d_s$ is the dimension of the global traffic state vector. Meanwhile, the MLP $k$ also outputs the corresponding immediate reward prediction,
\begin{align}
   R_{\theta_k}\left(\hat{s}(t),a(t),\hat{s}(t+1)\right),
\end{align}
where $R_{\theta_k}$ denotes the reward prediction head.

To enhance the diversity among different MLPs and prevent all models from converging to similar solutions, which may invalidate uncertainty estimation, we introduce a bootstrap-based data perturbation mechanism. At each parameter update, a mini-batch of size $B$, denoted as $\left\{\left(o^{(b)}(t),a^{(b)}(t),o^{(b)}(t+1)\right)\right\}_{b=1}^{B}$, is randomly sampled from the offline dataset $\mathcal{D}$, and the corresponding global traffic states are inferred using the state inference module, where $b$ denotes the sample index in the mini-batch. For the MLP $k$ and the training sample $b$, an independent Bernoulli mask is sampled as,
\begin{align}
    m_k^{(b)}\sim \mathrm{Bern}(p), \quad b=1,\ldots,B,\quad k=1,\ldots,K,
\end{align}
where $p$ denotes the sample retention probability. When $m_k^{(b)}=1$, the MLP $k$ uses this sample for parameter updating; otherwise, when $m_k^{(b)}=0$, this sample is excluded from gradient computation for the MLP $k$. This mechanism allows different MLPs to observe different subsets of data during each training iteration, thereby inducing diverse generalization behaviors in the learned state transition functions.

For learning the probabilistic state transition distribution of each MLP, we adopt the negative log-likelihood (NLL) loss as the training objective. Let the target state be $y=\mathrm{sg}\left(\hat{s}(t+1)\right)$, where $\mathrm{sg}(\cdot)$ denotes the stop-gradient operation. The state prediction loss for a single sample is defined as,
\begin{align}
    \mathcal{L}_{\mathrm{dyn}}=\frac{1}{2}\sum_{d=1}^{d_s}\left[\frac{\left(y^d-\mu_{\theta_k}^d\right)^2}{\left(\sigma_{\theta_k}^d\right)^2}+\log \left(\sigma_{\theta_k}^d\right)^2\right]+\frac{d_s}{2}\log(2\pi),
\end{align}
where $\mu_{\theta_k}^d$ and $\sigma_{\theta_k}^d$ denote the $d$-th elements of the predicted mean and standard deviation vectors, respectively.
For reward prediction, we use the mean squared error loss,
\begin{align}
\mathcal{L}_{\mathrm{rew}}=\left\|\hat{r}_{\theta_k}(t)-r(t)\right\|^2,
\end{align}
where $r(t)$ is the truth reward from the offline dataset. The training loss of the MLP $k$ is then defined as
\begin{align}
\mathcal{L}_k=\mathcal{L}_{\mathrm{dyn}}+\lambda_r \mathcal{L}_{\mathrm{rew}},
\end{align}
where $\lambda_r$ is the weighting coefficient for the reward prediction loss.
For a given state–action pair, the ensemble transition loss is formulated as,
\begin{align}
\mathcal{L}_{\mathrm{transition}}=\frac{\sum_{k=1}^{K}\sum_{b=1}^{B}\left(m_k^{(b)}\mathcal{L}_k^{(b)}\right)}{\sum_{k=1}^{K}\sum_{b=1}^{B}m_k^{(b)}}.
\end{align}

We integrate the global traffic state inference module and the probabilistic dynamics model into a unified traffic world model. The overall training objective is defined as,
\begin{align}
\mathcal{L}_{\mathrm{WM}}=\gamma_1 \mathcal{L}_{\mathrm{state}}+\gamma_2 \mathcal{L}_{\mathrm{transition}},
\end{align}
where $\gamma_1>0$ and $\gamma_2>0$ are weighting coefficients.

In addition, we characterize model epistemic uncertainty using the dispersion among the predicted means of different ensemble members,
\begin{align}
   & \chi\left(\hat{s}(t),a(t)\right)=\sqrt{\frac{1}{K}\sum_{k=1}^{K}{\|\mu_{\theta_k}\left(\hat{s}(t),a(t)\right)-\overline{\mu}\left(\hat{s}(t),a(t)\right)\|}^2},\\
   & \overline{\mu}\left(\hat{s}(t),a(t)\right)=\frac{1}{K}\sum_{k=1}^{K}\mu_{\theta_k}\left(\hat{s}(t),a(t)\right).
\end{align}
When a region is sufficiently covered by the offline dataset, the predictions of different ensemble members are generally consistent, resulting in low uncertainty. In contrast, when the policy visits state–action regions that are rarely covered by the training data, the predictions of different members may diverge significantly, leading to increased uncertainty. Therefore, $\chi\left(\hat{s}(t),a(t)\right)$ can serve as an important indicator of model extrapolation risk and is used in subsequent policy learning to restrict unreliable imagined interactions.

\subsection{World Model-Based Offline Multi-Agent Policy Learning}

After training, the world model is fixed and used to perform multi-step imagination rollouts to generate virtual trajectories for policy optimization. We adopt the CTDE paradigm and use multi-agent proximal policy optimization (MAPPO) \cite{yu2022surprising} to optimize the cooperative control policy for multiple CAVs. Specifically, during training, the recovered global traffic state and joint actions are used to evaluate the system-level return, thereby capturing the coupling effects among different CAV control behaviors. During execution, each CAV makes decisions independently based only on its local observations, satisfying the requirement for decentralized control in practical traffic scenarios.

However, policy optimization in world model-based offline learning is prone to model exploitation, where the policy exploits poorly covered state–action regions with overestimated returns, leading to spurious improvement.
To mitigate this issue, we construct a pessimistic reward based on the epistemic uncertainty estimated by the ensemble model,
\begin{align}
    \tilde{r}(\hat{s}(t),a(t))=\hat{r}(\hat{s}(t),a(t))-\beta \chi(\hat{s}(t),a(t)),
\end{align}
where $\beta$ is the pessimistic penalty coefficient. When the ensemble models exhibit large prediction disagreement for a given state–action pair, the pessimistic reward automatically reduces the estimated return in that region.

During imagination rollouts, we start from the recovered state $\hat{s}(t)$ sampled from the offline dataset and first use the observation decoder $O_{\psi}$ to obtain the local observation $\hat{o}_i(t)$ for each CAV. Each CAV then selects its control action according to its decentralized execution policy $\pi_{\nu_i}(a_i \mid \tau_i(t))$, and the individual actions are combined into the joint action $a(t)$. subsequently, the world model predicts the next state $\hat{s}(t+1)$ and outputs the pessimistic reward $\tilde{r}(\hat{s}(t),a(t))$ and epistemic uncertainty $\chi(\hat{s}(t),a(t))$. This procedure is repeated recursively to construct multi-step virtual trajectories.

To prevent error accumulation in long-horizon imagination rollouts and avoid policy learning from highly uncertain virtual experiences, we further introduce an epistemic uncertainty-based rollout truncation mechanism. Let $\chi_{\max}$ denote the threshold. If $\chi(\hat{s}(t),a(t))  \geq \chi_{\max}$ during rollouts, the current virtual trajectory is immediately terminated.

\subsection{Solution Algorithm}

Algorithm~\ref{alg:unified_framework} summarizes the overall training procedure of the proposed framework. The world model is parameterized by $\Theta=\{\phi,\psi,\xi,\theta_{1:K}\}$, where $\phi$, $\psi$, $\xi$, and $\theta_{1:K}$ denote the learnable parameters of the history encoder, observation decoder, cell-wise Cross-Attention module, and $K$ ensemble transition-and-reward models, respectively.  %6p
 \section{Numerical Results}
\label{sec_results}
\subsection{Experimental Setup}
\begin{figure}
    \raggedright
    \includegraphics[width=0.8\linewidth]{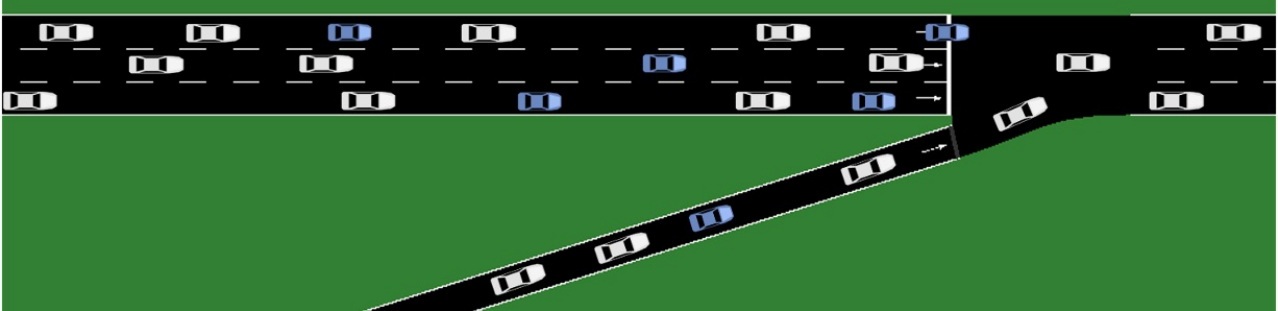}
    \caption{Simulation scenario of the on-ramp merging bottleneck}
    \label{fig:ramp_scenario}
\end{figure}

To evaluate the effectiveness of the proposed method, we conduct numerical experiments using the simulation of urban mobility (SUMO) platform in an on-ramp merging bottleneck scenario based on a section of the Hujin expressway, as illustrated in Fig.~\ref{fig:ramp_scenario}.
The CAV cooperative control zones on the mainline and on-ramp are $3000$~m and $600$~m long, respectively. 
Both are discretized with a spatial step of $\Delta x=300~\mathrm{m}$, resulting in $10$ traffic cells on the mainline and $2$ cells on the on-ramp.
The traffic demands on the mainline and on-ramp are set to $2500$~veh/h and $1800$~veh/h, respectively. The baseline CAV penetration rate is set to $15\%$, and the free-flow speed of all vehicles is $30$~m/s. The longitudinal acceleration of each CAV is constrained within $a_i\in[-3~m/s^2,3~m/s^2]$. Each simulation lasts for $1200$ time steps, with the first $200$ time steps used as a warm-up period.

\begin{algorithm}[!t]
\caption{Physics-Informed World Model-Based Offline MARL}
\label{alg:unified_framework}
\begin{algorithmic}[1]
\Require Offline dataset $\mathcal D$; ensemble size $K$; uncertainty threshold $\chi_{\max}$; pessimism coefficient $\beta$.
\Ensure World model $\Theta^*$ and decentralized policies $\Pi^*$.

\State Initialize world model $\Theta=\{\phi,\psi,\xi,\theta_{1:K}\}$
\State Initialize actors $\{\pi_{\nu_i}\}_{i\in\mathcal N}$ and critic $Q_{\varsigma}$
% \Statex \textbf{Phase I: Physics-Informed World Model Training}
\For{$iter=1,\ldots,N_{\mathrm{WM}}$}
    \State Sample mini-batch from $\mathcal D$
    \State Build histories $\{\tau_i(t),\tau_i(t+1)\}_{i\in\mathcal N}$
    \State Infer states $\hat s(t)$ and $\hat s(t+1)$
    \State Compute physical prior $s^{\mathrm{phy}}(t+1)=F_{\mathrm{phy}}(\hat s(t),a(t))$
    \State Predict $(\hat{s}_{\theta_k}(t+1),\hat{r}_{\theta_k}(t)) = MLP_{\theta_k}(\hat{s}(t),a(t)),\quad k=1,\ldots,K$
    \State Estimate uncertainty $\chi(t)$
    \State Update $\Theta$ by minimizing $L_{\mathrm{WM}}$
\EndFor
\State Set $\Theta^*\leftarrow\Theta$
\For{$iter=1,\ldots,N_{\mathrm{RL}}$}
    \State Sample history from $\mathcal D$ and infer initial state $\hat s(t_0)$
    \For{$h=0,\ldots,H_{\mathrm{img}}-1$}
        \State Decode observations and build $\{\hat\tau_i(t_0+h)\}_{i\in\mathcal N}$
        \State Sample actions $a_i(t_0+h)\sim\pi_{\nu_i}(\cdot\mid\hat\tau_i(t_0+h))$
        \State Form joint action $a(t_0+h)=\{a_i(t_0+h)\}_{i\in\mathcal N}$
        \State Predict $(\hat s(t_0+h+1),\hat r(t_0+h),\chi(t_0+h))$ using $\Theta^*$
        \If{$\chi(t_0+h)\geq\chi_{\max}$}
            \State \textbf{break}
        \EndIf
        \State Store transition with $\tilde r(t_0+h)$
    \EndFor
    \State Update $Q_{\varsigma}$ and $\{\pi_{\nu_i}\}_{i\in\mathcal N}$
\EndFor

\State \Return $\Theta^*$ and $\Pi^*=\{\pi_{\nu_i}^*\}_{i\in\mathcal N}$
\end{algorithmic}
\end{algorithm}

The offline dataset is generated using predefined behavior policies with different exploration levels under randomized traffic conditions. A total of $1000$ independent traffic trajectories are collected, resulting in approximately $1\times10^6$ state-transition samples. To avoid information leakage caused by temporal correlations between consecutive samples, the dataset is split at the trajectory level into training, validation, and test sets with a ratio of $70\%/15\%/15\%$. All offline learning methods use the same training data, local observations, and action spaces to ensure a fair comparison. 

The world model consists of $7$ independently initialized ensemble members. The historical observation window is set to $H=10$, and the imagination horizon is set to $H_{\rm img}=5$. The learning rates of the world model and policy networks are set to $1\times10^{-3}$ and $1\times10^{-4}$, respectively. The pessimistic reward coefficient is set to $\beta=0.5$, while the uncertainty truncation threshold $\chi_{\max}$ is determined as the $95$th percentile of the epistemic uncertainty measured on the validation set. All control experiments are independently conducted using $5$ random seeds, and the mean and standard deviation are reported.

\subsection{Baselines and Evaluation Metrics}

In addition to the without control case and the online multi-agent proximal policy optimization (MAPPO) \cite{yu2022surprising} baseline, we consider eight methods to systematically investigate the effects of physics supervision, global state representation, and world model-based learning on offline multi-agent cooperative control.

\begin{itemize}

    \item M1: Proposed. The proposed framework.

    \item {M2: WM-Physical.} This variant adopts the same physically interpretable state representation as M1 but removes the physics supervision from the state reconstruction process. Both the state reconstruction module and the world model are therefore trained in a fully data-driven manner.

    \item {M3: WM-Latent.}  This variant learns a global latent state without explicit physical interpretation from the local historical information of multiple CAVs and trains a purely data-driven world model in the resulting latent space.

    \item {M4: WM-Concat.} This variant directly concatenates the local observations of all CAVs to form the global information representation and trains a purely data-driven world model based on this representation.

    \item {M5-M8: Model-Free Offline Multi-agent Reinforcement Learning.} These methods directly learn control policies from the offline dataset without using a world model to generate imagined trajectories. To ensure a fair comparison, all four methods adopt the same multi-agent implicit q-learning (MAIQL)  \cite{tampuu2015multiagent} policy learning framework and differ only in their global state representations. Specifically, M5 uses the physics-supervised global physical state, M6 uses the global physical state reconstructed without physics supervision, M7 uses the learned latent state, and M8 uses the concatenation of all CAV local observations.

\end{itemize}

These baselines enable the contribution of each component to be examined from complementary perspectives. The comparison between M1 and M2 evaluates the effect of physics supervision. The paired comparisons M1/M5, M2/M6, M3/M7, and M4/M8 assess the benefit of introducing a world model under comparable global state representations. In addition, comparisons among M1-M4 and among M5-M8 are used to examine the influence of different global state representations.

The methods are evaluated from two aspects: world model prediction accuracy and traffic control performance. For the world model, the mean absolute error (MAE) of state prediction and reward prediction is used to evaluate predictive accuracy. For traffic control, average travel time (ATT), average waiting time (AWT), and average time loss (ATL) are adopted to evaluate the system-level traffic control performance of different methods.

\subsection{World-Model Prediction Accuracy}

We first evaluate the ability of different world models to predict traffic-state evolution and system rewards. As shown in Table~\ref{tab:wm_accuracy}, M1-M4 achieve good next-state prediction accuracy within their respective representation spaces. In particular, M4 uses the concatenation of all CAV local observations as the state representation, without requiring additional state reconstruction or latent feature mapping. Its prediction task is therefore relatively direct within its own representation space, resulting in a comparatively low state prediction error.
\begin{table}[h]
\raggedright
\caption{World-model prediction accuracy under different state representations}
\label{tab:wm_accuracy}
\small
\setlength{\tabcolsep}{4.0pt}
\renewcommand{\arraystretch}{1.15}
\begin{tabular}{lcccc}
\hline
Method 
& \makecell{Recovered-State\\MAE} 
& \makecell{Recovered-Reward\\MAE} 
& \makecell{True-State\\MAE} 
& \makecell{True-Reward\\MAE} \\
\hline
M1: Proposed    & 0.01633 & 0.02031  & 0.09543 & 0.02065 \\
M2: WM-Physical & 0.01912 & 0.02377  & 0.41197 & 0.02340 \\
M3: WM-Latent   & 0.06158 & 0.03221  & --      & --      \\
M4: WM-Concat   & 0.00704 & 0.03427  & --      & --      \\
\hline
\end{tabular}
\end{table}

\begin{figure}[h]
    \raggedright
    \begin{minipage}[t]{0.24\textwidth}
        \centering
        \includegraphics[width=\linewidth]{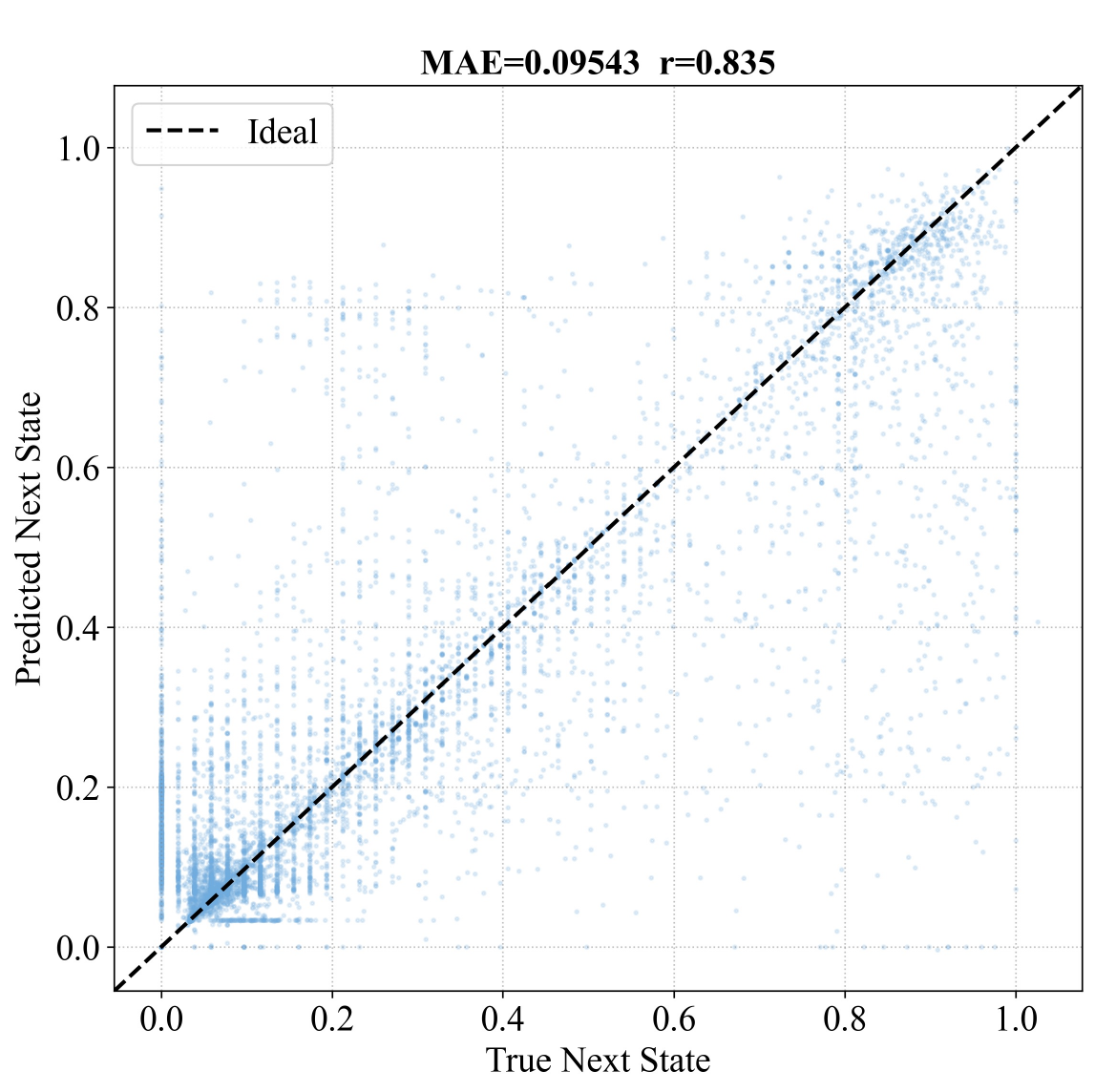}
        \small (a) M1: State
    \end{minipage}
    \hfill
    \begin{minipage}[t]{0.24\textwidth}
        \centering
        \includegraphics[width=\linewidth]{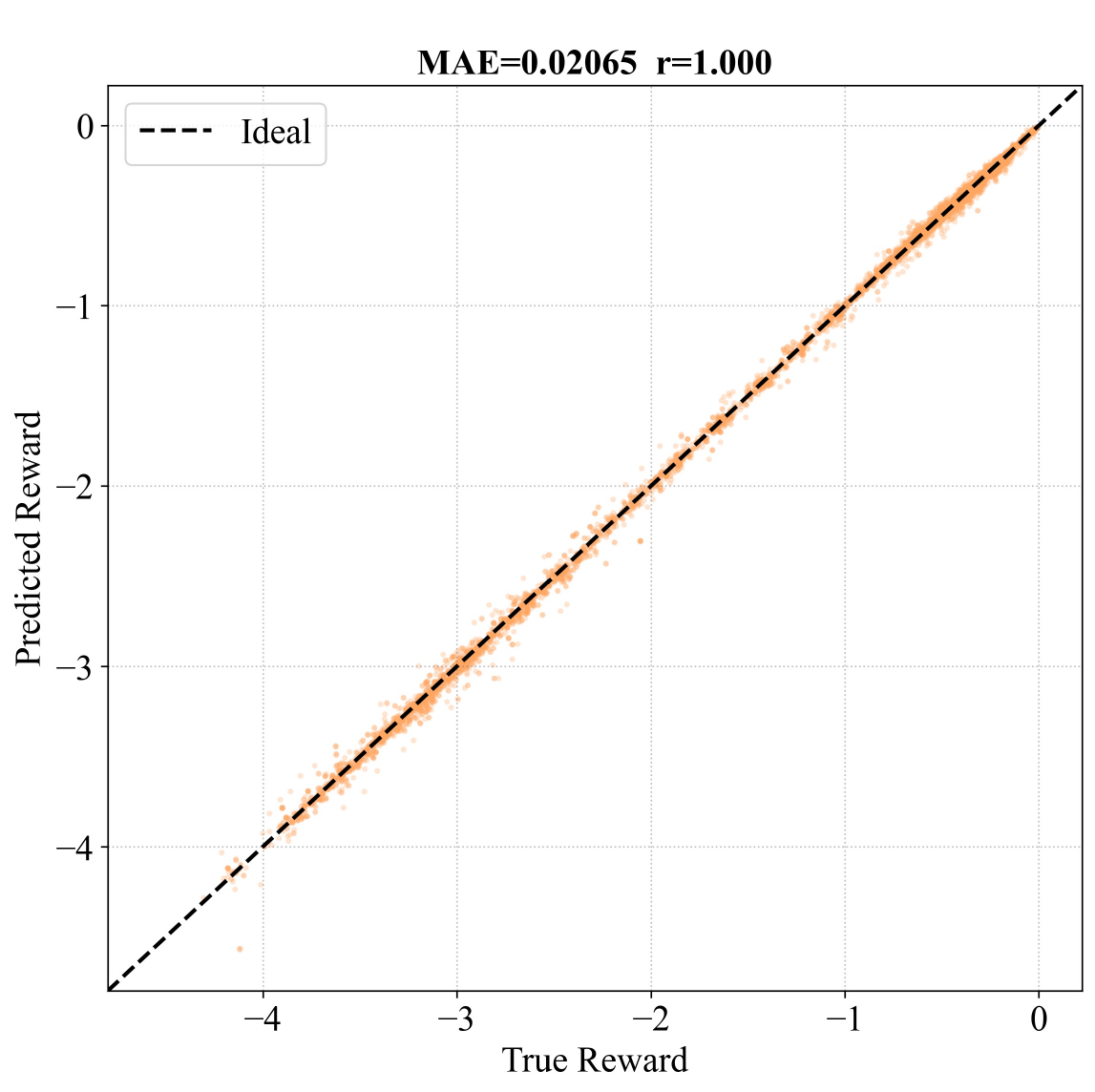}
        \small (b) M1: Reward
    \end{minipage}
    \hfill
    \begin{minipage}[t]{0.24\textwidth}
        \centering
        \includegraphics[width=\linewidth]{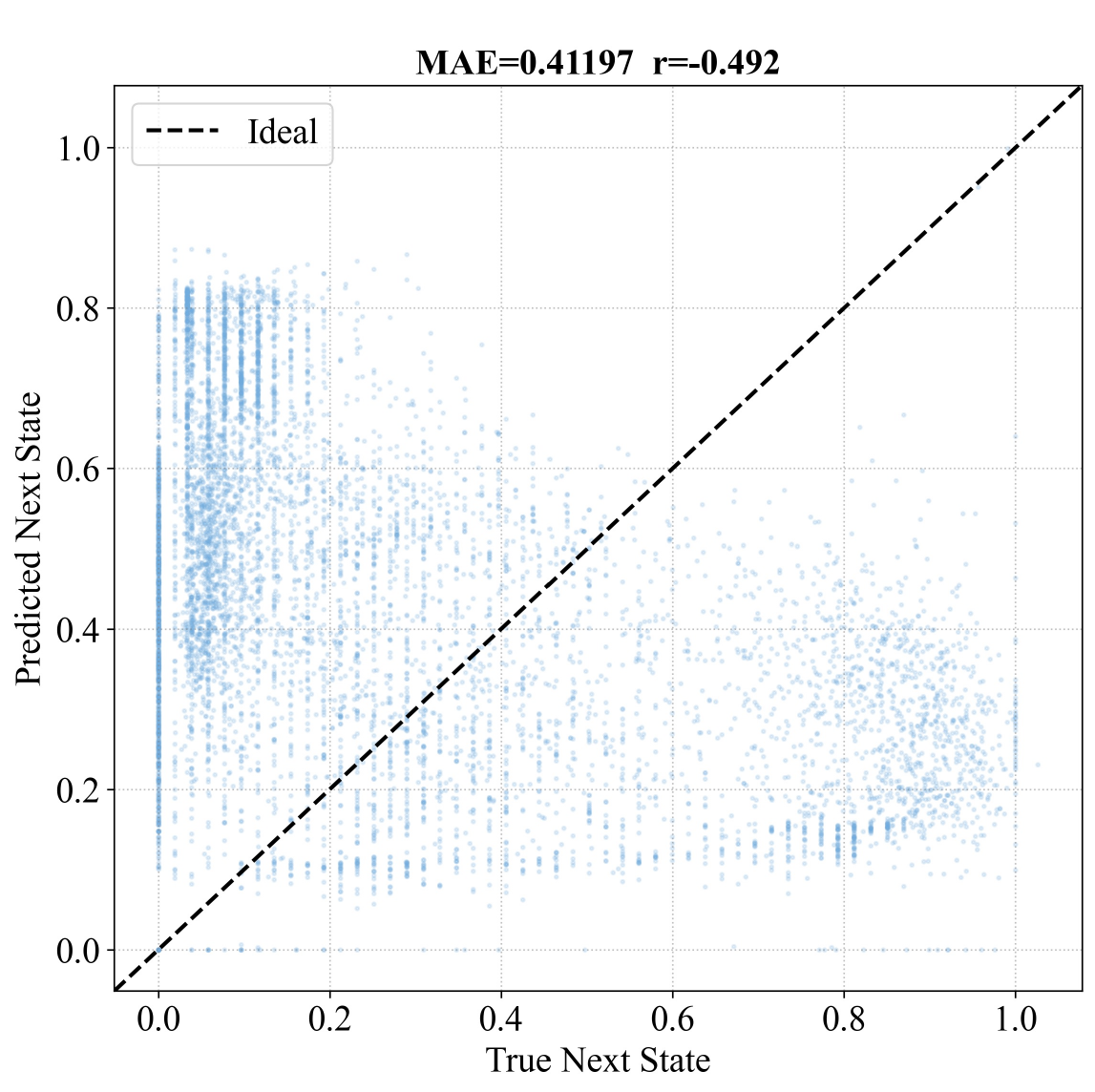}
        \small (c) M2: State
    \end{minipage}
    \hfill
    \begin{minipage}[t]{0.24\textwidth}
        \centering
        \includegraphics[width=\linewidth]{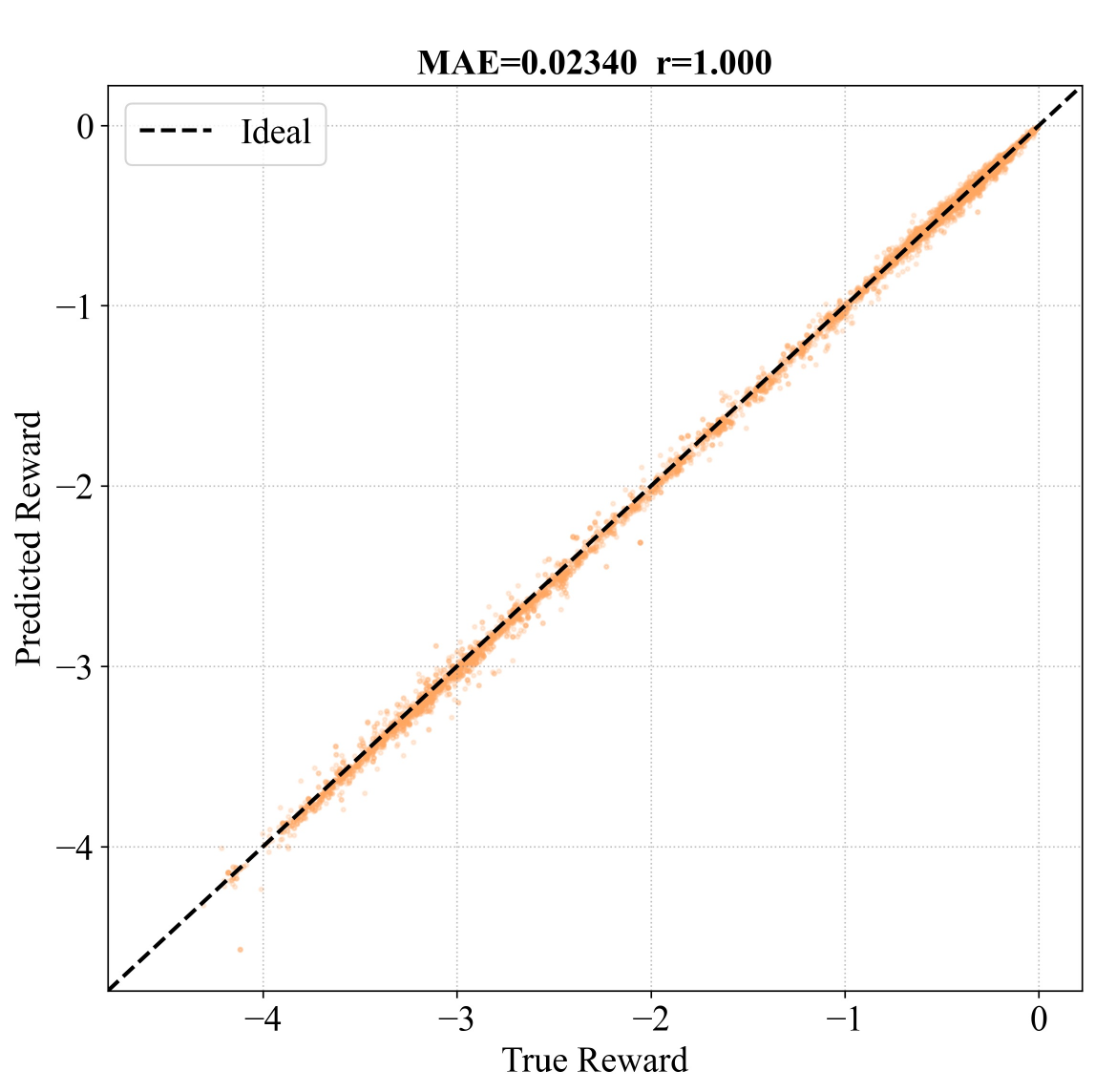}
        \small (d) M2: Reward
    \end{minipage}
    \caption{World-model prediction results of M1 and M2 against the SUMO ground truth}
    \label{fig:wm_prediction}
\end{figure}

For M1 and M2, whose state representations have explicit physical interpretations, we further compare their predictions with the ground-truth traffic states obtained from SUMO, as illustrated in Fig.~\ref{fig:wm_prediction}. The true-state MAE of M1 is $0.09543$, compared with $0.41197$ for M2, corresponding to a reduction of $76.84\%$. Similarly, the true-reward MAE decreases from $0.02340$ for M2 to $0.02065$ for M1, representing an improvement of $11.75\%$. These results indicate that traffic-flow physics guidance not only improves the predictive consistency within the reconstructed state space, but, more importantly, strengthens the correspondence between the reconstructed states and the actual traffic evolution. 

\subsection{Cooperative Control Performance}

Table~\ref{tab:control_performance} reports the traffic control performance of different methods in the on-ramp merging scenario shown in Fig.~\ref{ramp}. The proposed M1 achieves the best overall control performance among all offline learning methods, with the lowest ATT and ATL. Compared with the without control case, M1 reduces ATT, AWT, and ATL by $19.79\%$, $61.62\%$, and $36.10\%$, respectively. Moreover, its performance approaches that of Online MAPPO, indicating that the proposed method can learn effective cooperative control policies for multiple CAVs without requiring online interaction with the environment.

\begin{table}[h]
\raggedright
\caption{Traffic control performance of different methods}
\label{tab:control_performance}
\small
\setlength{\tabcolsep}{4.2pt}
\renewcommand{\arraystretch}{1.15}
\begin{tabular}{lcccccccccc}
\hline
Metric 
& \makecell{Without\\Control} & \makecell{Online\\MAPPO} & M1 & M2 & M3 & M4 & M5 & M6 & M7 & M8 \\
\hline
ATT (s) & 145.81 & 107.93 & \textbf{116.96} & 122.92 & 127.59 & 132.14 & 120.69 & 127.40 & 129.23 & 135.90 \\
AWT (s) & 0.99 & 0.14 & 0.38 & \textbf{0.32} & 0.37 & 0.34 & 0.38 & 0.53 & 0.51 & 0.49 \\
ATL (s) & 85.44 & 47.51 & \textbf{54.60} & 65.43 & 68.33 & 76.32 & 57.10 & 70.37 & 72.80 & 81.21 \\
\hline
\end{tabular}
\end{table}

By incorporating physics supervision, M1 reduces ATT and ATL by $4.85\%$ and $16.55\%$, respectively, compared with M2, although its AWT increases slightly. Combined with the world-model prediction results in Section~4.3, these results further suggest that physics supervision improves the reliability of global state reconstruction and world-model prediction, which in turn leads to better overall system-level traffic control performance.

In addition, the paired comparisons between M1-M4 and their corresponding model-free counterparts M5-M8 show that world model-based policy learning generally achieves better traffic control performance under comparable state representations. This indicates that imagined rollouts generated by the learned world model can effectively expand the coverage of the fixed offline dataset and provide additional training experience for offline policy optimization, thereby improving the learned cooperative control policies.

\subsection{Sensitivity Analysis}

To further evaluate the sensitivity of the proposed method under different traffic conditions, we examine the effects of traffic demand and CAV penetration rate on control performance.

\begin{figure}[h]
    \raggedright
    \begin{minipage}[t]{0.32\textwidth}
        \centering
        \includegraphics[width=\linewidth]{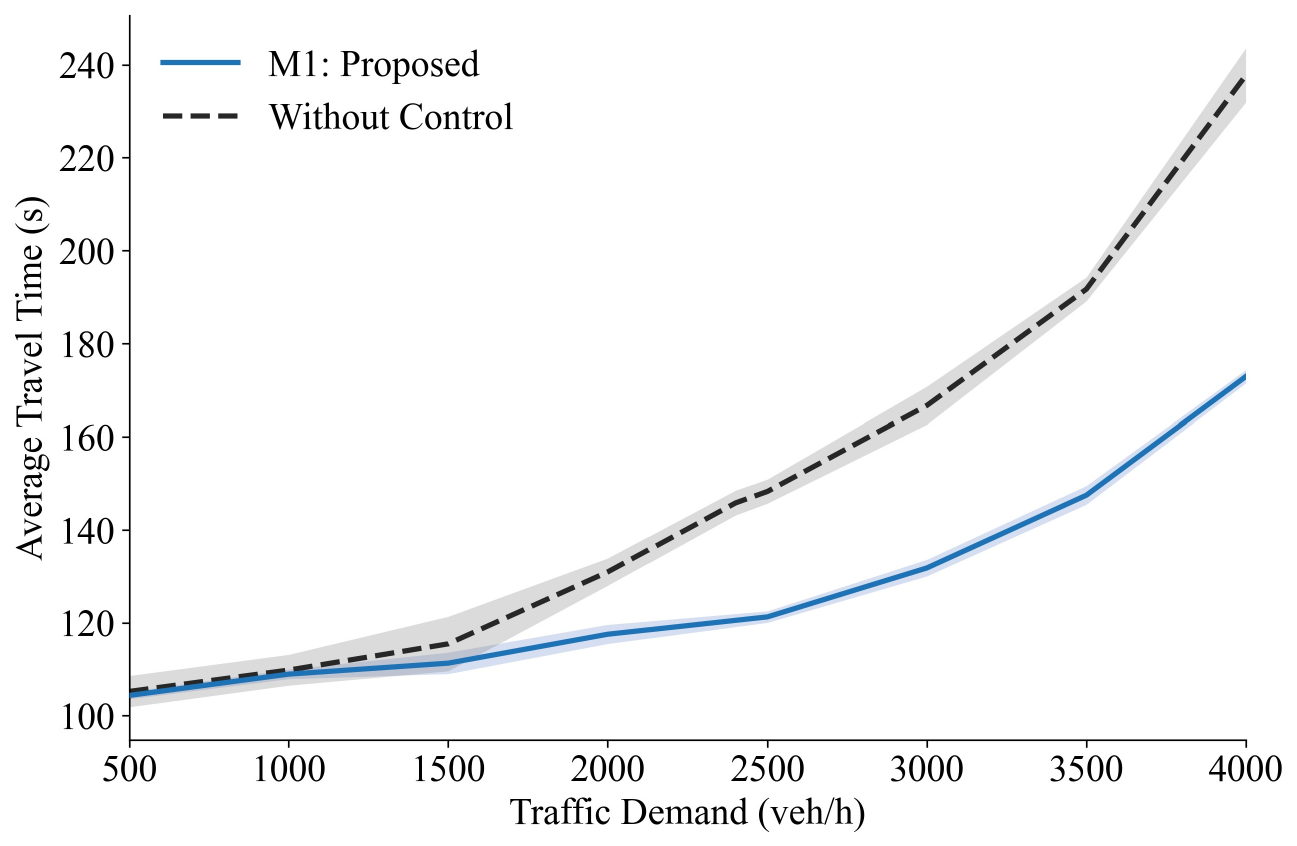}
        \vspace{-2mm}
        \centerline{\small (a) Average travel time}
    \end{minipage}
    \hfill
    \begin{minipage}[t]{0.32\textwidth}
        \centering
        \includegraphics[width=\linewidth]{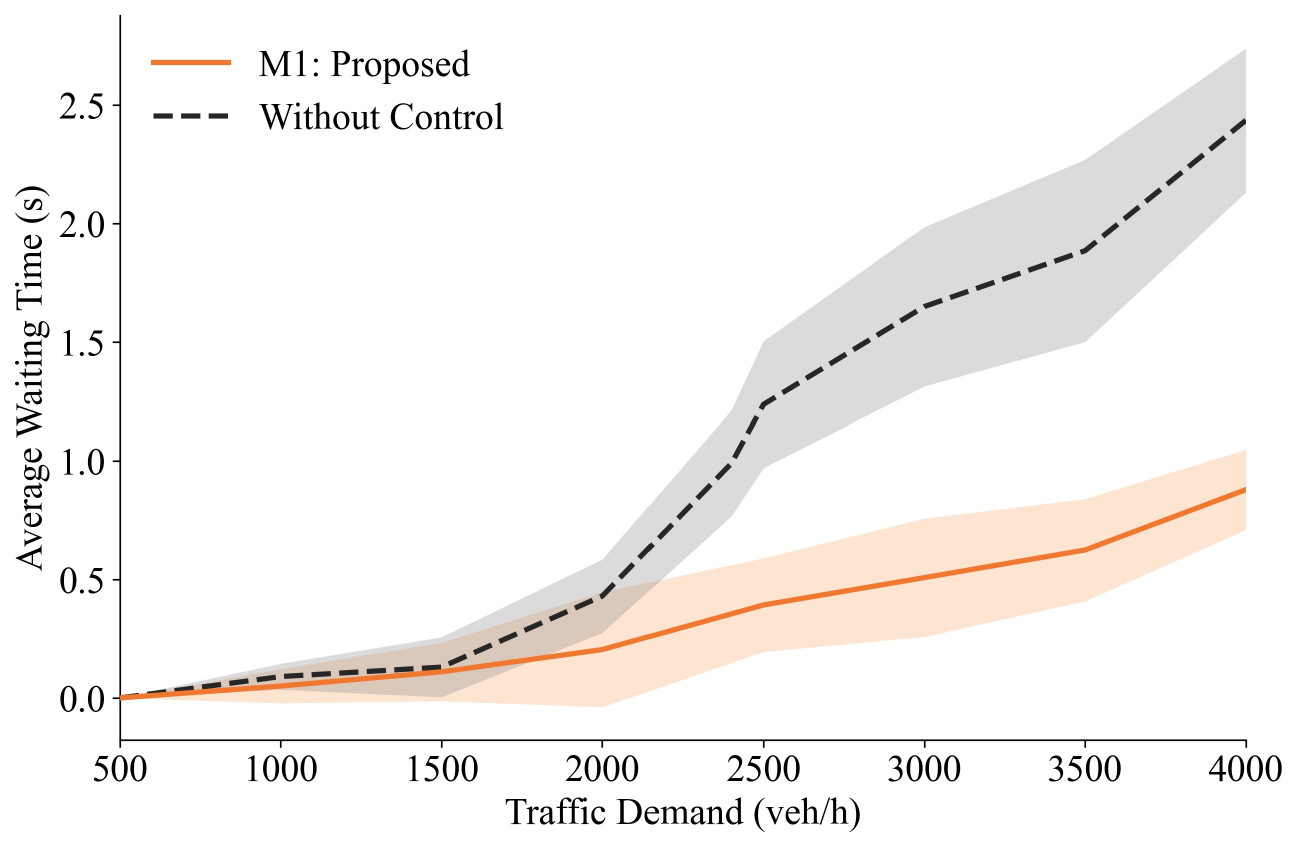}
        \vspace{-2mm}
        \centerline{\small (b) Average waiting time}
    \end{minipage}
    \hfill
    \begin{minipage}[t]{0.32\textwidth}
        \centering
        \includegraphics[width=\linewidth]{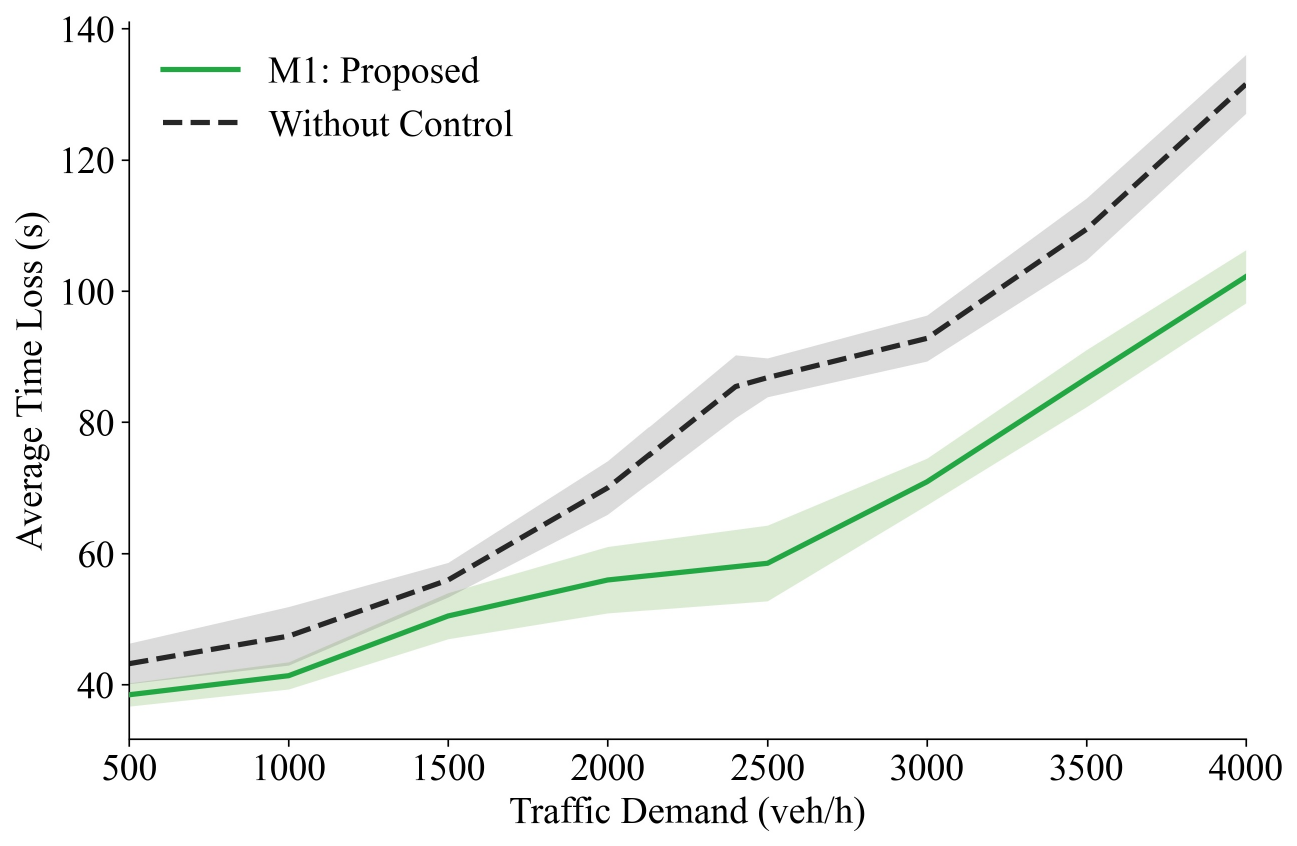}
        \vspace{-2mm}
        \centerline{\small (c) Average time loss}
    \end{minipage}
    \caption{Sensitivity of traffic control performance to traffic demand}
    \label{fig:demand_sensitivity}
\end{figure}

Fig.~\ref{fig:demand_sensitivity} compares the traffic performance of the proposed method with the without control case under different traffic demand levels. Under low traffic demand, the bottleneck operates under relatively uncongested conditions, resulting in only a small difference in ATT between the two cases. As traffic demand increases and bottleneck congestion intensifies, the performance gap progressively widens.
In particular, when traffic demand exceeds approximately $2000~\mathrm{veh/h}$, ATT, AWT, and ATL increase rapidly in the without control case, whereas the proposed method effectively suppresses their growth and maintains effective congestion mitigation even under high-demand conditions.

\begin{figure}[h]
    \raggedright
    \begin{minipage}[t]{0.32\textwidth}
        \centering
        \includegraphics[width=\linewidth]{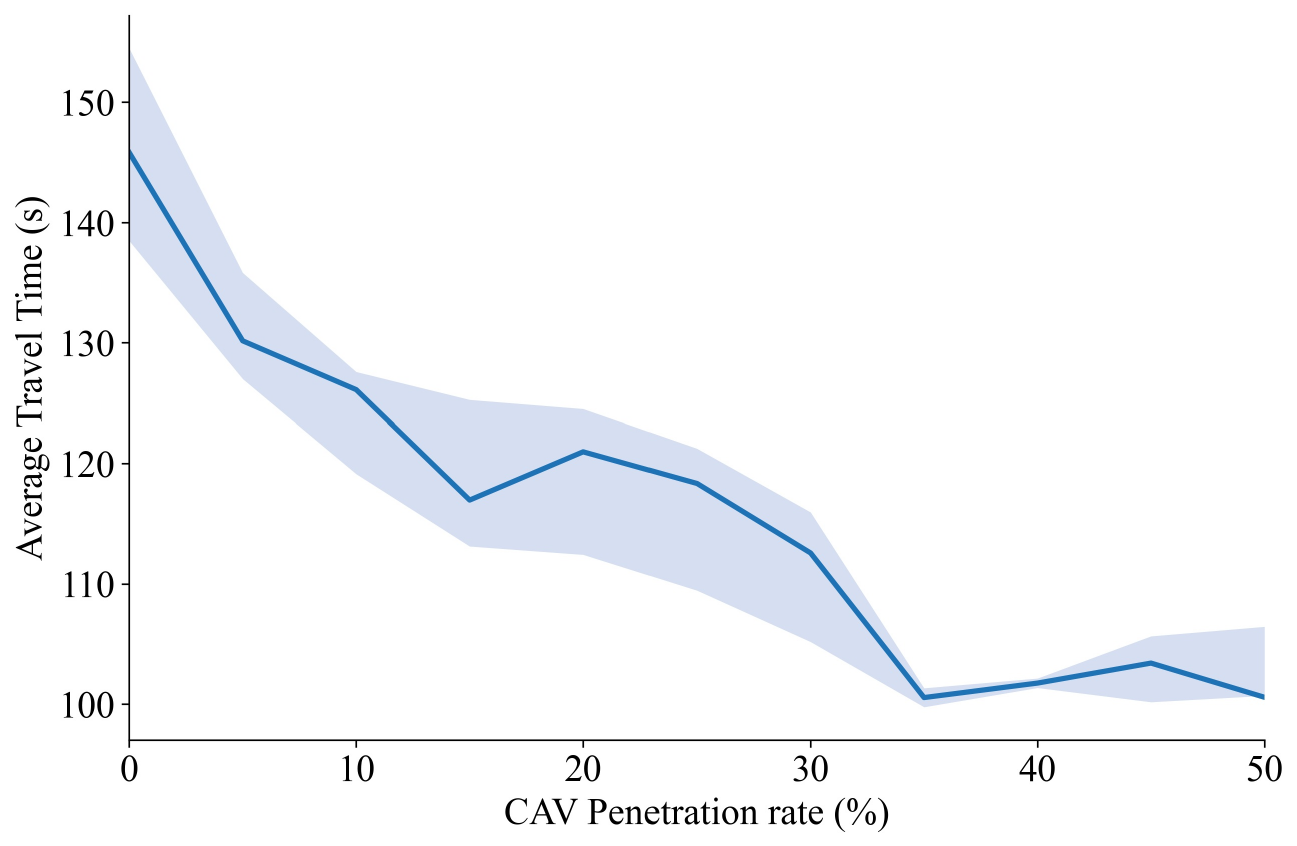}
        \vspace{-2mm}
        \centerline{\small (a) Average travel time}
    \end{minipage}
    \hfill
    \begin{minipage}[t]{0.32\textwidth}
        \centering
        \includegraphics[width=\linewidth]{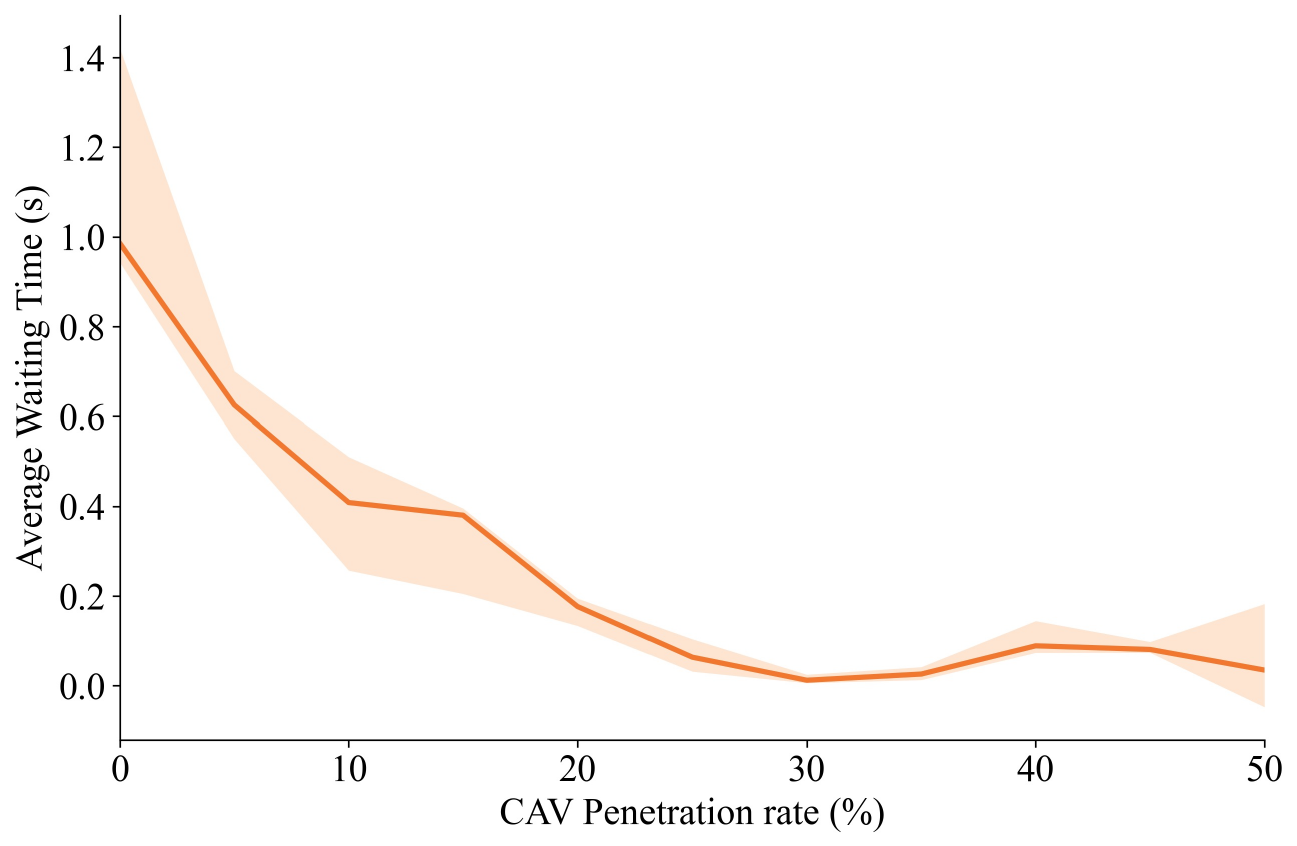}
        \vspace{-2mm}
        \centerline{\small (b) Average waiting time}
    \end{minipage}
    \hfill
    \begin{minipage}[t]{0.32\textwidth}
        \centering
        \includegraphics[width=\linewidth]{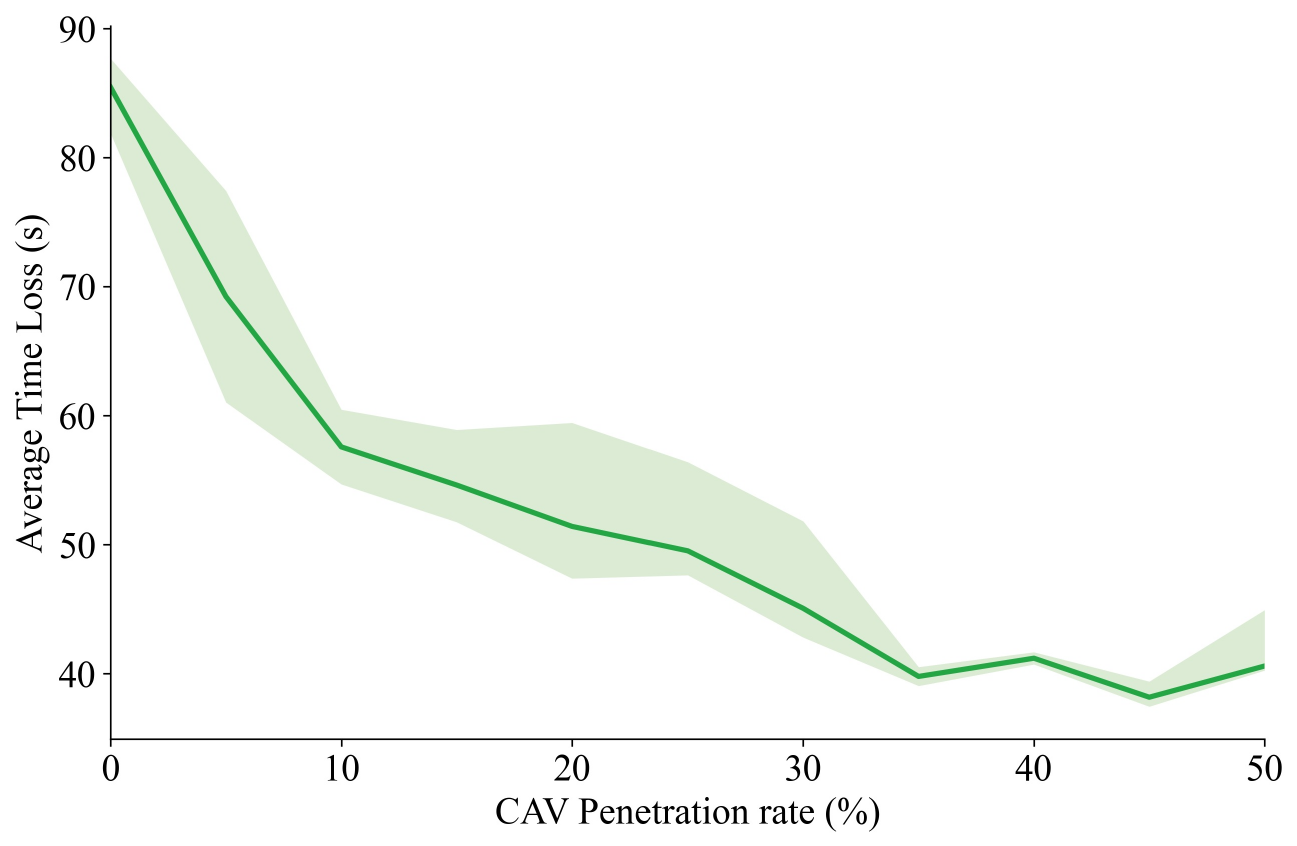}
        \vspace{-2mm}
        \centerline{\small (c) Average time loss}
    \end{minipage}
    \caption{Sensitivity of traffic control performance to CAV penetration rate}
    \label{fig:cav_sensitivity}
\end{figure}

Fig.~\ref{fig:cav_sensitivity} further investigates the effect of CAV penetration rate. Overall, ATT, AWT, and ATL exhibit decreasing trends as the CAV penetration rate increases, with the most pronounced improvements occurring at relatively low penetration levels. When the penetration rate reaches approximately $30\%$-$35\%$, the marginal benefits of further increasing the number of CAVs gradually diminish. This indicates that once a sufficient number of CAVs are available to effectively regulate upstream traffic flow, additional controllable vehicles provide progressively smaller performance gains.
 %3p
 \section{Conclusions}
\label{sec_conclude}

This paper proposes a physics-informed world model-based offline multi-agent reinforcement learning framework for cooperative CAV control at partially observable mixed-traffic bottlenecks. The proposed method reconstructs a structured global traffic state with explicit physical meaning from the local observation-action histories of multiple CAVs and enhances its consistency with traffic evolution through a coupled macroscopic-microscopic traffic model. Based on the reconstructed state, a probabilistic ensemble world model is developed to learn traffic state transitions and system rewards. Pessimistic rewards and uncertainty-driven rollout truncation are further incorporated to generate reliable imagined trajectories for offline cooperative policy learning.
Simulation experiments based on a real-world on-ramp bottleneck demonstrate that physics supervision improves the reliability of state reconstruction and world-model prediction, leading to better cooperative control performance. Sensitivity analysis further shows that the proposed method remains effective under varying traffic demands and CAV penetration rates and provides consistent congestion mitigation under congested conditions. Overall, the proposed framework enables effective cooperative CAV control without online trial-and-error interactions or access to complete global traffic states.

Future work will focus on validating the proposed framework using field-collected trajectory and CAV observation data, investigating its robustness to sensing and communication uncertainties as well as traffic-model mismatch, and extending the framework to more complex bottleneck configurations and larger-scale traffic networks. These extensions will further assess the practical transferability and generalization capability of the proposed method under realistic traffic conditions. %1p

% ---------- Acknowledgments ----------
\section*{Acknowledgments}
The authors used OpenAI ChatGPT to assist with language editing and the identification of potentially relevant literature. All technical content and research results were independently developed by the authors, and all references were independently verified. The authors take full responsibility for the content of this manuscript.

% ---------- Author Contributions ----------
% Consider using the Contributor Role Taxonomy (CRediT) https://credit.niso.org/ when stating author contributions
\section*{Author Contributions}
The authors confirm contribution to the paper as follows:
study conception and methodology: Lu Liu, Xi Xiong;
data collection, experiments, and manuscript preparation: Lu Liu;
manuscript review and revision: Lu Liu, Chi Xie, Xi Xiong.
All authors reviewed and approved the final version of the manuscript.

% ---------- Declaration of COI ----------
\section*{Declaration of Conflicting Interests}
% Choose one of the following statements:
The authors declared no potential conflicts of interest with respect to the research, authorship, and/or publication of this article.

% % OR
%The authors declared the following potential conflicts of interest with respect to the research, authorship, and/or publication of this article: \emph{[insert text here]}.

% % OR
% The authors declared no potential conflicts of interest with respect to the research, authorship, and/or publication of this article.

\section*{Funding}
% Choose one of the following statements:
The authors disclosed no financial support for the research, authorship, and/or publication of this article.

% % OR
% The authors disclosed no financial support for the research, authorship, and/or publication of this article.

\bibliographystyle{unsrtnat}
\bibliography{trb_template}

@article{Richards1956ShockWO,
  title={Shock Waves on the Highway},
  author={Paul I. Richards},
  journal={Operations Research},
  year={1956},
  volume={4},
  number={1},
  pages={42-51}
}

@article{DAGANZO1994269,
title = {The Cell Transmission Model: A Dynamic Representation of Highway Traffic Consistent with the Hydrodynamic Theory},
journal = {Transportation Research Part B: Methodological},
volume = {28},
number = {4},
pages = {269-287},
year = {1994},
author = {Carlos F. Daganzo}
}

@article{JIN20121000PartB,
title = {A Kinematic Wave Theory of Multi-Commodity Network Traffic Flow},
journal = {Transportation Research Part B: Methodological},
volume = {46},
number = {8},
pages = {1000-1022},
year = {2012},
author = {Wen-Long Jin}
}

@INPROCEEDINGS{Mladen2018Traffic,   
author={Čičić, Mladen and Johansson, Karl Henrik},   
booktitle={2018 21st International Conference on Intelligent Transportation Systems},    
title={Traffic Regulation via Individually Controlled Automated Vehicles: {A} Cell Transmission Model Approach},    
year={2018},   
volume={},   
number={},   
pages={766-771} 
}

@ARTICLE{Xiong2024ITS,
  author={Xiong, Xi and Wang, Maonan and Sun, Dengfeng and Jin, Li},
  journal={IEEE Transactions on Intelligent Transportation Systems}, 
  title={An Approximate Dynamic Programming Approach to Vehicle Platooning Coordination in Networks}, 
  year={2024},
  volume={25},
  number={11},
  pages={16536-16547}
  }

@ARTICLE{Feng2021,
  author={Feng, Shuo and Song, Ziyou and Li, Zhaojian and Zhang, Yi and Li, Li},
  journal={IEEE Transactions on Intelligent Vehicles}, 
  title={Robust Platoon Control in Mixed Traffic Flow Based on Tube Model Predictive Control}, 
  year={2021},
  volume={6},
  number={4},
  pages={711-722}
  }

@article{LI2020225,
title = {Trajectory Data-Based Traffic Flow Studies: {A} Revisit},
journal = {Transportation Research Part C: Emerging Technologies},
volume = {114},
pages = {225-240},
year = {2020},
issn = {0968-090X},
author = {Li Li and Rui Jiang and Zhengbing He and Xiqun Chen and Xuesong Zhou}
}

@article{WANG2023359,
title = {Hierarchical Attention Master–Slave for Heterogeneous Multi-Agent Reinforcement Learning},
journal = {Neural Networks},
volume = {162},
pages = {359-368},
year = {2023},
issn = {0893-6080},
author = {Jiao Wang and Mingrui Yuan and Yun Li and Zihui Zhao}
}

@article{Zhou2025aaai,
author = {Zhou, Yihe and Zheng, Yuxuan and Hu, Yue and Chen, Kaixuan and Zheng, Tongya and Song, Jie and Song, Mingli and Liu, Shunyu},
year = {2025},
month = {04},
pages = {23018-23026},
title = {Cooperative Policy Agreement: {Learning} Diverse Policy for Offline MARL},
volume = {39},
journal = {Proceedings of the AAAI Conference on Artificial Intelligence},
doi = {10.1609/aaai.v39i21.34465}
}

@inproceedings{Ross2021ICML,
author = {Ross, Stephane and Bagnell, J. Andrew},
year = {2012},
month = {03},
pages = {},
title = {Agnostic System Identification for Model-Based Reinforcement Learning},
volume = {2},
booktitle = {Proceedings of the 29th International Conference on Machine Learning, ICML 2012}
}

@inproceedings{Kidambi2020NIPS,
 author = {Kidambi, Rahul and Rajeswaran, Aravind and Netrapalli, Praneeth and Joachims, Thorsten},
 booktitle = {Advances in Neural Information Processing Systems},
 pages = {21810--21823},
 title = {MOReL: {Model-Based} Offline Reinforcement Learning},
 volume = {33},
 year = {2020}
}

@inproceedings{Kang2025ICLR,
 author = {Kang, Sehyeok and Lee, Yongsik and Kim, Gahee and Chong, Song and Yun, Se-Young},
 booktitle = {International Conference on Learning Representations},
 pages = {20145--20165},
 title = {{M}$\text{A}^2${E}: {Addressing} Partial Observability in Multi-Agent Reinforcement Learning with Masked Auto-Encoder},
 volume = {2025},
 year = {2025}
}

@ARTICLE{Liao2022ITS,
  author={Liao, Xishun and Wang, Ziran and Zhao, Xuanpeng and Han, Kyungtae and Tiwari, Prashant and Barth, Matthew J. and Wu, Guoyuan},
  journal={IEEE Transactions on Intelligent Transportation Systems}, 
  title={Cooperative Ramp Merging Design and Field Implementation: {A} Digital Twin Approach Based on Vehicle-to-Cloud Communication}, 
  year={2022},
  volume={23},
  number={5},
  pages={4490-4500}
}

@article{QIU2023Cooperative,
    title = {Cooperative Trajectory Control for Synchronizing the Movement of Two Connected and Autonomous Vehicles Separated in a Mixed Traffic Flow},
    journal = {Transportation Research Part B: Methodological},
    volume = {174},
    pages = {102769},
    year = {2023},
    author = {Jiahua Qiu and Lili Du}
}

@INPROCEEDINGS{Fang2022ITSC,
  author={Fang, Xing and Zhang, Qichao and Gao, Yinfeng and Zhao, Dongbin},
  booktitle={2022 IEEE 25th International Conference on Intelligent Transportation Systems}, 
  title={Offline Reinforcement Learning for Autonomous Driving with Real World Driving Data}, 
  year={2022},
  volume={},
  number={},
  pages={3417-3422}
  }

@InProceedings{pmlr-v202-ran23a,
  title = {Policy Regularization with Dataset Constraint for Offline Reinforcement Learning},
  author = {Ran, Yuhang and Li, Yi-Chen and Zhang, Fuxiang and Zhang, Zongzhang and Yu, Yang},
  booktitle = {Proceedings of the 40th International Conference on Machine Learning},
  pages =  {28701--28717},
  year = {2023},
  volume = {202}
}

@article{cang2021behavioral,
  title={Behavioral Priors and Dynamics Models: Improving Performance and Domain Transfer in Offline {RL}},
  author={Cang, Catherine and Rajeswaran, Aravind and Abbeel, Pieter and Laskin, Michael},
  journal={arXiv preprint arXiv:2106.09119},
  year={2021}
}

@article{li2023developing,
  title={Developing a Dynamic Speed Control System for Mixed Traffic Flow to Reduce Collision Risks near Freeway Bottlenecks},
  author={Li, Ye and Pan, Bin and Chen, Zhibin and Xing, Lu},
  journal={IEEE Transactions on Intelligent Transportation Systems},
  volume={24},
  number={11},
  pages={12560--12581},
  year={2023},
  publisher={IEEE}
}

@article{huang2025uncertainty,
  title={Uncertainty-Based Alternative Diffusion Policy for Safe Autonomous Driving},
  author={Xiaohan Huang and Xuesong Wang and Yuhu Cheng},
  journal={IEEE Transactions on Intelligent Transportation Systems},
  year={2025},
  volume={26},
  pages={18854-18863}
}

@article{wiesemann2013robust,
  title={Robust Markov Decision Processes},
  author={Wiesemann, Wolfram and Kuhn, Daniel and Rustem, Ber{\c{c}}},
  journal={Mathematics of Operations Research},
  volume={38},
  number={1},
  pages={153--183},
  year={2013},
  publisher={INFORMS}
}

@article{yu2022surprising,
  title={The Surprising Effectiveness of {PPO} in Cooperative Multi-Agent Games},
  author={Yu, Chao and Velu, Akash and Vinitsky, Eugene and Gao, Jiaxuan and Wang, Yu and Bayen, Alexandre and Wu, Yi},
  journal={Advances in neural information processing systems},
  volume={35},
  pages={24611--24624},
  year={2022}
}

@article{rauch2024cooperative,
  title={Cooperative Multi-Agent Deep Reinforcement Learning for Dynamic Task Execution and Resource Allocation in Vehicular Edge Computing},
  author={Rauch, Robert and Becvar, Zdenek and Mach, Pavel and Gazda, Juraj},
  journal={IEEE Transactions on Vehicular Technology},
  volume={74},
  number={4},
  pages={5741--5756},
  year={2024},
  publisher={IEEE}
}

@article{hafner2019dream,
  title={Dream to Control: Learning Behaviors by Latent Imagination},
  author={Hafner, Danijar and Lillicrap, Timothy and Ba, Jimmy and Norouzi, Mohammad},
  journal={arXiv preprint arXiv:1912.01603},
  year={2019}
}

@article{bui2024comadice,
  title={{ComaDICE}: Offline Cooperative Multi-Agent Reinforcement Learning with Stationary Distribution Shift Regularization},
  author={Bui, The Viet and Nguyen, Thanh Hong and Mai, Tien},
  journal={arXiv preprint arXiv:2410.01954},
  year={2024}
}

@article{bernstein2002complexity,
  title={The Complexity of Decentralized Control of {M}arkov Decision Processes},
  author={Bernstein, Daniel S and Givan, Robert and Immerman, Neil and Zilberstein, Shlomo},
  journal={Mathematics of operations research},
  volume={27},
  number={4},
  pages={819--840},
  year={2002},
  publisher={INFORMS}
}

@article{newell1993simplified,
  title={A Simplified Theory of Kinematic Waves in Highway Traffic, Part II: {Queueing} at Freeway Bottlenecks},
  author={Newell, Gordon F},
  journal={Transportation Research Part B: Methodological},
  volume={27},
  number={4},
  pages={289--303},
  year={1993},
  publisher={Elsevier}
}

@article{LU202226partb,
title = {Are Autonomous Vehicles Better Off Without Signals at Intersections? {A} Comparative Computational Study},
journal = {Transportation Research Part B: Methodological},
volume = {155},
pages = {26-46},
year = {2022},
author = {Gongyuan Lu and Zili Shen and Xiaobo Liu and Yu Nie and Zhiqiang Xiong}
}

@article{CHUNG200782PartB,
title = {Relation Between Traffic Density and Capacity Drop at Three Freeway Bottlenecks},
journal = {Transportation Research Part B: Methodological},
volume = {41},
number = {1},
pages = {82-95},
year = {2007},
author = {Koohong Chung and Jittichai Rudjanakanoknad and Michael J. Cassidy}
}

@article{tampuu2015multiagent,
  title={MultiAgent Cooperation and Competition with Deep Reinforcement Learning},
  author={Tampuu, Ardi and Matiisen, Tambet and Kodelja, Dorian and Kuzovkin, Ilya and Korjus, Kristjan and Aru, Juhan and Aru, Jaan and Vicente, Raul},
  journal={arXiv preprint arXiv:1511.08779},
  year={2015}
}
\end{document}